\documentclass{INTERSPEECH2023}
\usepackage{graphicx}
\usepackage{float}
\interspeechcameraready

\title{Enhancing Speech Articulation Analysis Using A Geometric Transformation of the X-ray Microbeam Dataset}
\name{Ahmed Adel Attia$^1$, Mark Tiede $^2$, Carol Y. Espy-Wilson$^1$\thanks{This work was supported by National Science Foundation grant 2141413}}
\address{
  $^1$University of Maryland College Park, MD, $^2$ Haskins Laboratories, New Haven, CT}
\email{aadel@umd.edu, mark.tiede@yale.edu, espy@umd.edu}

\begin{document}

\maketitle
 
\begin{abstract}
Accurate analysis of speech articulation is crucial for speech analysis. However, X-Y coordinates of articulators strongly depend on the anatomy of the speakers and variability of pellet placements, and existing methods for mapping anatomical landmarks in the X-ray Microbeam Dataset (XRMB) fail to capture the entire anatomy of the vocal tract. In this paper, we propose a new geometric transformation that improves the accuracy of these measurements. Our transformation maps anatomical landmarks’ X-Y coordinates along the midsagittal plane onto six relative measures: Lip Aperture (LA), Lip Protrusion (LP), Tongue Body Constriction Location (TBCL), Degree (TBCD), Tongue Tip Constriction Location (TTCL), and Degree (TTCD). Our novel contribution is the extension of the palate trace towards the inferred anterior pharyngeal line, which improves measurements of tongue body constriction. 
 
\end{abstract}

\section{Introduction}

Articulatory data provides valuable insights into speech production and analysis. By recording the movement and position of articulators such as the lips, tongue, and jaw during speech, researchers can gain a better understanding of how speech sounds are produced. Articulatory data are used in a wide variety of applications like Automatic Speech Recognition \cite{tvasr, tvasr2}, speech synthesis \cite{speech_synthesis_1}, speech therapy \cite{Fagel2008A3V}, and mental health assessment \cite{mentalhealth, mentalhelth_2}. There are several methods for collecting articulatory data, including the University of Wisconsin X-ray Microbeam,  Electromagnetic Articulometry (EMA), and real-time Magnetic Resonance Imaging (rt-MRI).

However, accurately analyzing articulatory data can be challenging due to speaker anatomy and pellet placement variability. The positioning of pellets in the X-Y plane is closely linked to the speaker's anatomy, leading to significant variability in pellet positions among speakers for the same sound. Additionally, small differences in pellet placement can result in considerable variation. Speech production involves creating constrictions at various locations along the vocal tract by shaping the vocal tract filter using the articulators. Furthermore, the absolute positions of the articulators depend on the speaker's vocal tract anatomy. Therefore, quantifying vocal tract shape is best achieved by measuring the location and degree of these constrictions, which are relative measures, rather than the absolute X-Y positions of the pellets. These measures are called Tract Variables (TVs). TVs specify the salient features of the vocal tract area function more directly than the
pellet trajectories \cite{MCGOWAN199419}.  

Geometric transformations can be used to derive TVs from the absolute X-Y pellet positions. These variables, introduced in \cite{BrowmanGoldstein}, provide information on the location and degree of constrictions in the vocal tract without requiring knowledge of the absolute positions of the articulators. The TV Lip Aperture (LA), for example, quantifies the degree of constriction at the lips without differentiating the contributions of the jaw, upper lip, or lower lip \cite{sivaraman2019unsupervised}.\\
In this paper, we focus on the XRMB dataset. \cite{sivaraman2019unsupervised} describes a geometric transformation to obtain TVs from the XRMB Pellet Trajectories (PTs). According to the Task Dynamic model of speech production (TADA) \cite{nam2004tada}, the hard palate and tongue body were approximated as circles using curve fitting techniques. Specifically, the hard palate was approximated as a large circle using curve fitting through the palate trace, and the tongue body was approximated as a smaller circle within the larger circle that approximates the palate. The tongue tip was modeled separately using the segment T2-T1. This modeling approach enabled the conversion of the pellet X-Y positions to six TV trajectories at each time step. These trajectories included LA, Lip Protrusion (LP), Tongue Body Constriction Location (TBCL), Tongue Body Constriction Degree (TBCD), Tongue Tip Constriction Location (TTCL), and Tongue Tip Constriction Degree (TTCD). This transformation has been proven to be effective in modeling articulations in the vocal tract and has been used in various publications \cite{siriwardena2022audio, espywilson19_interspeech, seneviratne20_interspeech, siriwardena2021multimodal}. Recently, \cite{attia2022masked} outlined an Autoencoder model to reconstruct 3.28 out of 3.4 hours of corrupted XRMB data. However, they reconstructed the X-Y pellet positions and applying \cite{sivaraman2019unsupervised}'s transformation to their reconstructed data proved to be difficult as only a high-level description of their transformations is outlined and no formulae or code was available. 

The transformation model described \cite{sivaraman2019unsupervised} also has several limitations. One major drawback is that the palatal trace in the XRMB dataset only covers a limited area and does not include the soft palate or extend to the pharyngeal wall. As a result, the model cannot accurately account for the role of these structures in speech production, and cannot model narrowings in the pharyngeal region.

Another limitation of the model is that the circular arc used to model the palatal trace may not accurately represent the actual shape of the palate. To overcome these limitations, our paper proposes a new approach that incorporates the soft palate and anterior pharyngeal wall traces into the model. This will improve the accuracy of the tongue body constriction calculation, especially in back vowels. Our approach also differs from the previous model in that it utilizes the palatal trace as is from the XRMB dataset, without the need to fit a circular arc through it. By addressing these limitations, our model will provide a more accurate representation of speech production. We also describe our transformation in more detail so that our work can be reproducible, and be applied to \cite{attia2022masked}'s reconstructed XRMB dataset.

\section{Dataset Description}
The XRMB dataset consists of simultaneously recorded audio and articulatory data. During the recording process, eight gold pellets were placed on specific articulators in each speaker, including the upper lip (UL), lower lip (LL), tongue tip (T1), tongue blade (T2), tongue dorsum (T3), tongue root (T4), mandible incisor (MNI), and parasagittally placed mandible molar (MNM). Speakers were given various tasks, such as reading passages or word lists, while their articulatory movements were tracked and recorded as X-Y coordinates. The data was sampled at different rates, so to ensure consistency, all pellet trajectories were resampled at a rate of 145 samples per second. Multi-sentence recordings were segmented into individual sentence recordings. In some cases, articulatory recordings were marked as mistracked in the database, reducing the dataset to 46 speakers (21 males and 25 females) with approximately 4 hours of speech data. 
\begin{figure*}[h!]
 \center

  \includegraphics[width=0.8\textwidth]{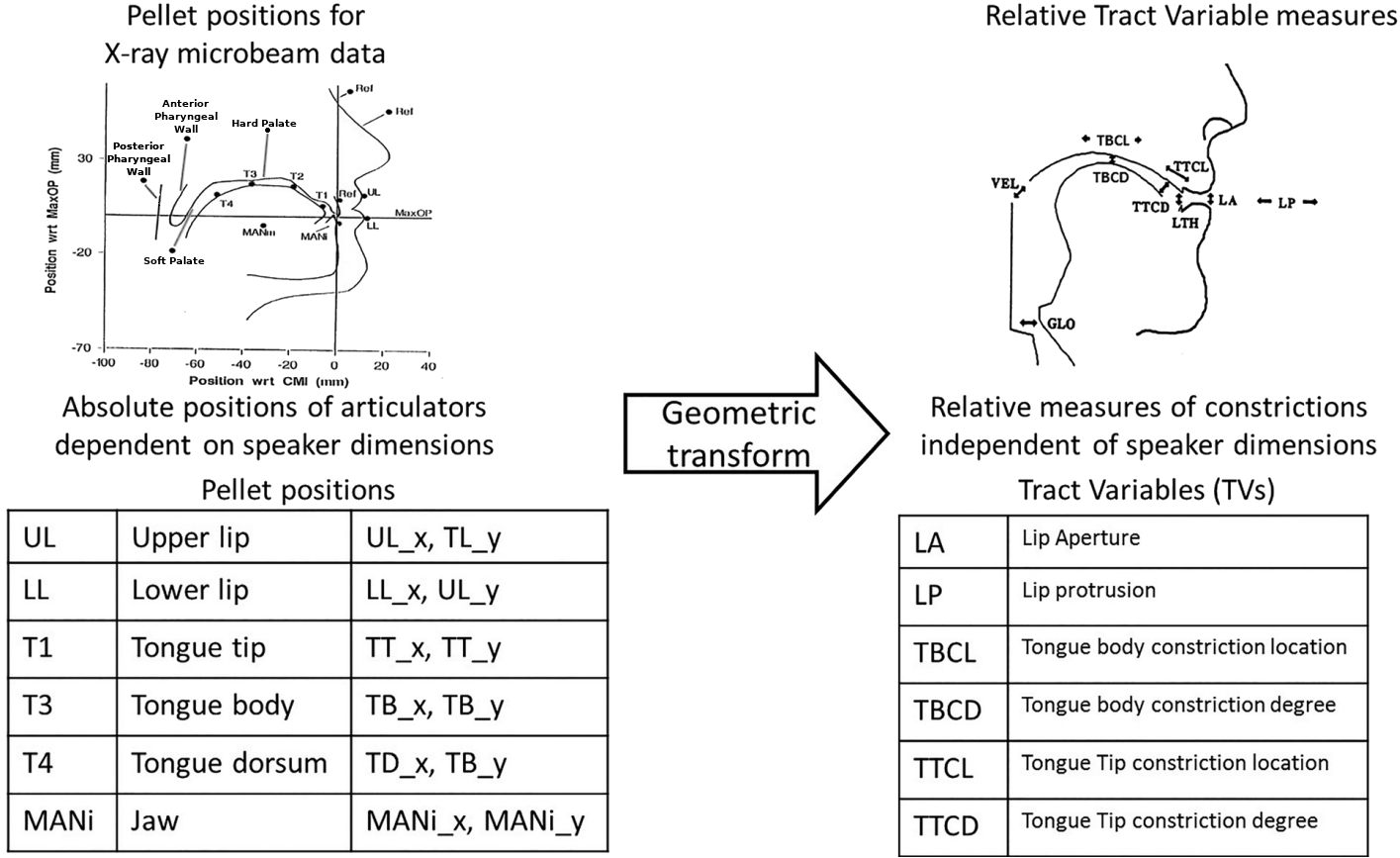}
  \caption{Transformation of XRMB PTs to TVs. Figure taken from \cite{sivaraman2019unsupervised}}
  \label{TV trans}
\end{figure*}
\section{Geometric Transformation}

In this section, we will provide a comprehensive overview of the proposed method for transforming the XRMB PTs into TVs, providing a detailed explanation of each articulator. 
\begin{figure}[H]
    \centering
    \includegraphics[width=\columnwidth]{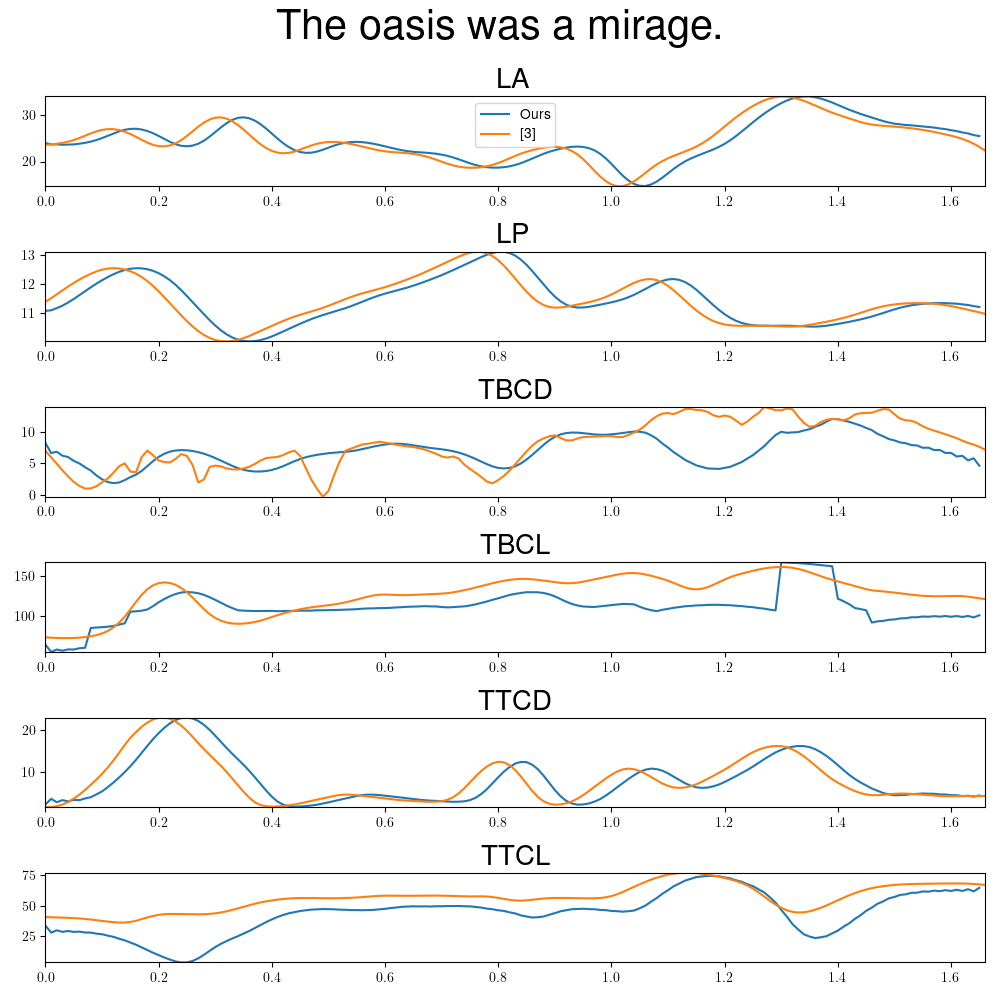}
    \caption{Our novel TVs vs \cite{sivaraman2019unsupervised} TVs. }
    \label{fig:my_label}
\end{figure}
\subsection{Lips}
To describe the process of lip constriction, we use two different TV values: LA and LP. LA represents the distance between the upper lip (UL) and lower lip (LL) pellets in Euclidean space, while LP represents the horizontal offset of the upper lip from the origin. 
It is important to note that the origin of the X-Y plane is located at the tip of the maxillary incisors, and the X-axis is defined as the maxillary occlusal plane. This means that any measurements or calculations made in relation to lip constriction are made with reference to this specific orientation. 
\begin{equation}
    LA[n] = || UL[n] - LL[n]||
\end{equation}
\vspace{-15pt}
\begin{equation}
    LP[n] = UL_x[n]
\end{equation}

\subsection{Tongue Body}
The tongue body is represented by a circle that is fitted through the pellets T2, T3, and T4. The degree of constriction in the tongue body is determined based on the proximity of this circle to the palate trace that is extended towards the anterior pharyngeal wall. This method allows for an accurate representation of tongue constriction during speech production.

In the XRMB dataset, posterior pharyngeal wall traces are available for all speakers. By shifting the posterior pharyngeal wall to the right by the thickness of the low retropalatal oropharynx, we can infer the trace of the anterior pharyngeal wall. The average low retropalatal oropharyngeal thickness has been estimated to be approximately 0.58 cm for women and 0.56 cm for men \cite{DANIEL20075}.

To estimate the position of the soft palate, we extend the line between the two posterior samples in the palate trace in the XRMB dataset until it intersects with the inferred anterior pharyngeal wall. 

\begin{figure}[H]
    \centering
\includegraphics[width=\columnwidth]{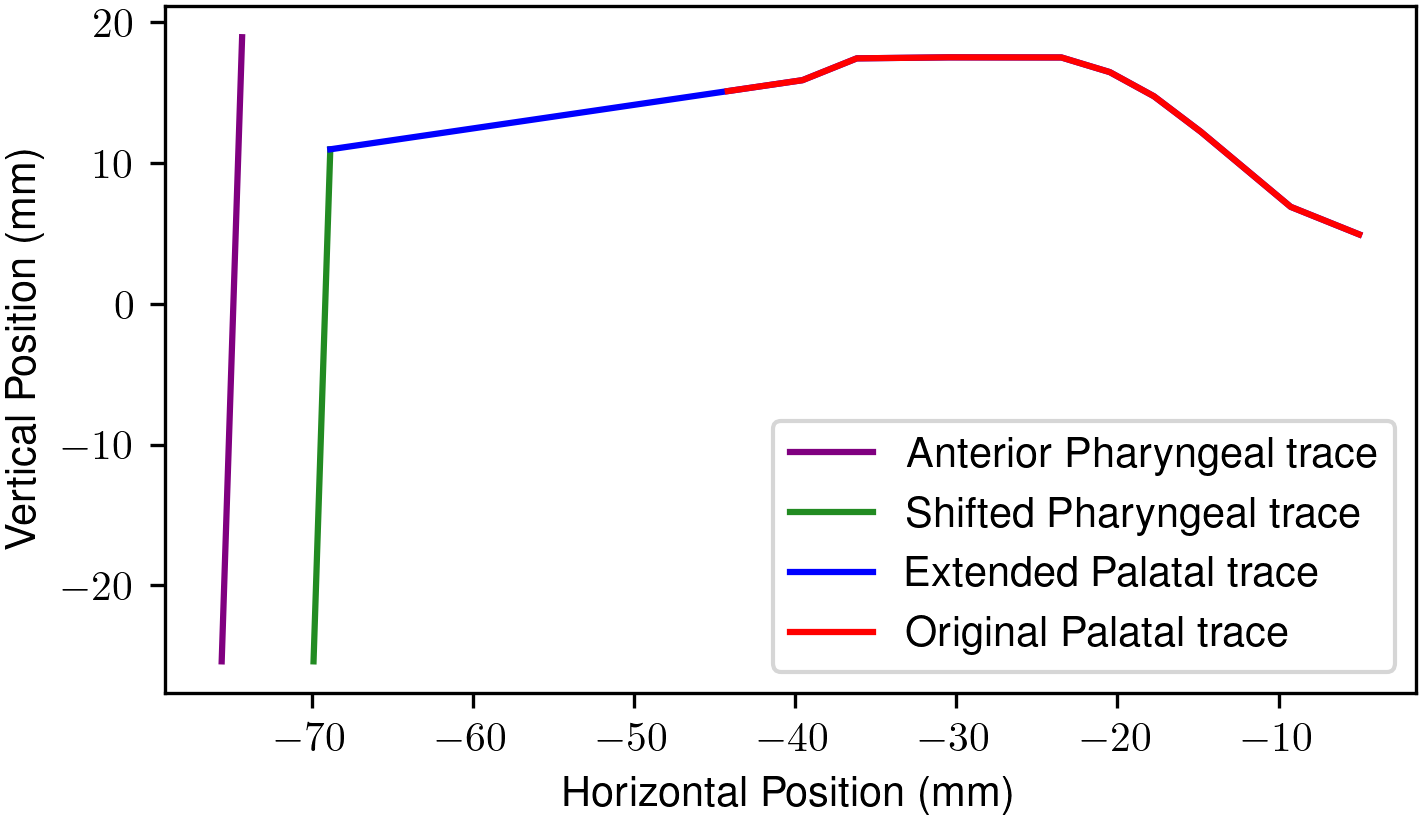}
    \caption{Extended Palateal Trace With the Anterior Pharyngeal Wall For Speaker JW33.}
\label{fig:extended_pal}
\end{figure}

Figure \ref{fig:extended_pal} illustrates the original and extended palatal trace, as well as the anterior and posterior pharyngeal wall traces, providing a visual representation of the methodology used in this process. By using these techniques, we can create precise computer models of speech production that accurately represent the positions of the tongue and soft palate.

The tongue body constriction is described by TBCD and TBCL. \cite{mitra2012recognizing} defines TBCL (and TTCL) as polar angular distance measures for the tongue body and tongue tip constrictors with respect to a reference line originating at the floor of the mouth. In our approach, we follow a similar methodology, but we use the center of a circle that is fitted through the original palatal trace as the reference point instead. It is important to note that this circle is not used in any calculations; only its center is utilized as a reference point for angular calculations. 

Therefore, TBCD can be calculated as the minimum distance between the extended palate trace and the tongue body circle while TBCL represents the angle between the center of the palatal circle and the point on the tongue body that is closest to the extended palate trace. Notice in Figure \ref{fig:my_label} how TBCD is less noisy than \cite{sivaraman2019unsupervised}'s TBCD and describes a narrower constriction around 1.3 seconds, which corresponds to the $\Lambda$ sound in the word "mirage". Our model was more capable of capturing this pharyngeal narrowing in the back vowel.

 \begin{equation}
     TBCD = min_{p \in epal}[ min_{x \in TB_{circle}} || p - x||]
 \end{equation}
 \begin{equation}
  TBCL = tan ^{-1} \frac{TB[argmin[TBCD]_x] - pc_x} {TB[argmin[TBCD]_y - pc_y}
 \end{equation}
 where $epal$ is the extended palate trace, $TB[argmin[TBCD]$ is the point on the tongue body closest to the palate trace, and $pc$ is the center of the palate circle.
\subsection{Tongue Tip}
The tongue tip is represented by the T1 pellet, and its constriction can be modeled by TTCD and TTCL. Similar to the tongue body, TTCD is the minimum distance between T1 and the extended palate trace, while TTCL is the horizontal angle of the segment between the center of the pallet circle and T1. 
 \begin{equation}
     TTCD = min_{p \in epal}[ || p - T1||]
 \end{equation}
 \begin{equation}
  TBCL = tan ^{-1} \frac{T1_x - pc_x} {T1_y - pc_y}
 \end{equation}
\section{Experiments}
We conducted experiments to evaluate our novel geometric transformation by training a Speech Inversion (SI) system on two sets of TVs: our own and the ones used in \cite{sivaraman2019unsupervised}. The Bidirectional Gated Recurrent Neural Network (BiGRNN) SI system from \cite{siriwardena2022audio} was used, along with Mel-Frequency Cepstral Coefficients (MFCCs) extracted from 2-second audio segments. Shorter segments were zero-padded. The dataset was divided into three sets: training (36 speakers), development (5 speakers), and testing (5 speakers, 3 males, and 2 females), with no overlapping speakers to ensure "speaker-independent" training. The split ensured that the training and development/testing sets contained around 80\% and nearly equal amounts of utterances, respectively. All models were built with the TensorFlow-Keras machine learning framework and trained on a NVIDIA Titan Xp GPU using MAE loss with the Adam optimizer. We also used early stopping with a patience of 10 epochs during the training process. Early stopping is a technique used to prevent overfitting in machine learning models by monitoring the validation loss during training. The training process is stopped when the validation loss stops improving after a certain number of epochs, called "patience". In our case, the training process would stop if the validation loss did not improve for 10 consecutive epochs. This technique helps to prevent the model from overfitting to the training data, which can lead to poor performance on new, unseen data. Both models were evaluated using the Pearson Product Moment Correlation (PPMC) score.

Table \ref{table: model_comp} shows the PPMC score for models trained with both sets of TVs. The SI system scores 6 points higher with TBCD showing that the inferred soft palate and the anterior pharyngeal wall  yield a measure of the tongue body constriction that is more correlated with the data. On average, our TVs score 3 points more than previous works.
\vspace*{-3pt}
\begin{table}[th]
\Large
  \caption{PPMC scores for models trained with our TVs and \cite{sivaraman2019unsupervised}'s TVs}
  \vspace*{-3pt}
  \centering
  \label{table: model_comp}
  \resizebox{\columnwidth}{!}
  {\begin{tabular}{|l|l|l|l|l|l|l|l|}
    \hline 
                \multicolumn{1}{|l|}{\textbf{TVs}}  &      \multicolumn{1}{|l|}{\textbf{LA}}  &  \multicolumn{1}{|l|}{\textbf{LP}}&  \multicolumn{1}{|l|}{\textbf{TBCL}}&  \multicolumn{1}{|l|}{\textbf{TBCD}}&  \multicolumn{1}{|l|}{\textbf{TTCL}}&  \multicolumn{1}{|l|}{\textbf{TTCD}}&  \multicolumn{1}{|l|}{\textbf{Average}} 
    \\[1em]\hline
    &&&&&&&\\
    \textbf{Ours} & \textbf{0.6587} & \textbf{0.6068} & 0.6455 & \textbf{0.6259} & \textbf{0.5621} & 0.6817 & \textbf{0.6301}\\[1em]
    \textbf{\cite{sivaraman2019unsupervised}} & 0.6239 & 0.5430 & \textbf{0.6823} & 0.5616 & 0.5271 & \textbf{0.6940} & 0.6054
    \\[1em]\hline
  \end{tabular}}
\end{table}
\section{Conclusions And Future Works}
We introduce a novel geometric transformation to derive the Tongue Body Constriction Location (TBCL) TVs from X-Y coordinates of articulatory trajectories in the XRMB dataset. Our approach involves incorporating an inferred outline of the soft palate and anterior pharyngeal wall, resulting in a more accurate measurement of tongue body constriction.

However, we acknowledge that our current model, which represents the tongue body as a circular arc, has room for improvement and could benefit from further investigation. We observed that our TBCL TV is not continuous, which could explain why the Speech Inversion (SI) system performed better on predicting the TBCL in \cite{sivaraman2019unsupervised}. In future research, we plan to address these limitations.
\bibliographystyle{IEEEtran}
\bibliography{mybib}

\end{document}